# Tsallis Relative entropy from asymmetric distributions as a risk measure for financial portfolios

Sandhya Devi[1] and Sherman Page[2]

Edmonds, WA, 98020, USA

Email: sdevi@entropicdynamics.com

sherman@entropicdynamics.com

**Abstract.** In an earlier study, we showed that Tsallis relative entropy (TRE), which is the generalization of Kullback-Leibler relative entropy (KLRE) to non-extensive systems, can be used as a possible risk measure in constructing risk optimal portfolios whose returns beat market returns. Over a long term ($> 10$ years), the risk-return profiles from TRE as the risk measure show a more consistent behavior than those from the commonly used risk measure 'beta' of the Capital Asset Pricing Model (CAPM). In these investigations, the model distributions derived from TRE are symmetric. However, observations show that distributions of the returns of financial markets and equities are in general asymmetric in positive and negative returns. In this work, we generalize TRE for the asymmetric case (ATRE) by considering the data distribution as a linear combination of two independent normalized distributions – one for negative returns and one for positive returns. Each of these two independent distributions are half $q$-Gaussians with different non-extensivity parameter $q$ and temperature parameter $b$. The risk-return (in excess of market returns) patterns are investigated using ATRE as the risk measure. The results are compared with those from two other risk measures: TRE and the Tsallis relative entropy $S_-$ derived from the negative returns only. Tests on data, which include the dot-com bubble, the 2008 crash, and COVID periods, for both long (~20 years) and shorter terms (~10 years), show that a linear fit can be obtained for the risk-excess return profiles of all three risk measures. However, the fits for portfolios created during the chaotic market conditions (crashes) using $S_-$ as the risk show a much higher slope pointing to higher returns for a given risk value. Further, in this case, the excess returns of even short-term portfolios remain positive irrespective of the market behavior.

---

[1] Shell International Exploration and Production Co. (Retired)
[2] Shell Information Technology International (Retired)

## 1. Introduction

In capital asset management, risk optimal portfolios are usually based on using the covariance $\beta$ defined in modern portfolio theory [1] and the capital asset pricing model (CAPM) [2][3][4][5][6] or simply the standard deviation $\sigma$ as volatility measures. These measures are based on the efficient market hypothesis [7][8] according to which a) investors have all the information available to them and they independently make rational decisions using this information, b) the market reacts to all the information available reaching equilibrium quickly, and c) in this equilibrium state the market has a normal distribution. Under these conditions, the return $R_j$ for an equity $j$ is linearly related to the market return $R_m$ [9] as

$$R_j = \beta_j R_m + \alpha_j + \hat{e}_j \tag{1a}$$

$\beta_j$ is the risk parameter given by

$$\beta_j = \rho_{j,m} (\sigma_j / \sigma_m) \tag{1b}$$

where $\rho_{j,m}$ is the correlation coefficient of $R_j$ and $R_m$, and $\sigma_j$ and $\sigma_m$ are the standard deviations of $R_j$ and $R_m$. $\hat{e}_j$ is a normally distributed error term.

The intercept $\alpha_j$ is the value of $R_j$ when $R_m$ is zero and hence can be considered as the excess return of the equity above the market return. The return $R$ over a period $\tau$ is defined as

$$R(t,\tau) = (X(t) - X(t - \tau))/X(t - \tau) \tag{1c}$$

$X(t)$ is the stock value at time $t$.

In 1972, empirical tests of the validity of CAPM were carried out by Black, Jensen and Scholes [10] who examined the monthly returns of all the stocks listed in NYSE for 35 years, between 1931-1965. Portfolios were constructed by binning the estimated risk parameter $\beta$ and allocating the stocks for each bin according to their risk parameter. The long term (35 years) results showed a highly linear relationship between the excess portfolio return $\alpha$ and the bin risk parameter $\beta$, the slope being slightly positive. This indicates that the higher risk stock portfolios yield marginally higher excess returns. However, when the tests were carried out for shorter periods (~9 years), the relationship between the excess returns and $\beta$ were still linear but the slopes were non-stationary becoming even negative for some periods.

In reality, how true are the assumptions of CAPM? Observations show that the market is a complex system that is the result of decisions by interacting agents (e.g., herding behavior), traders who speculate and/or act impulsively on little news, etc. Such a collective/chaotic behavior can lead to wild swings in the system, driving it away from equilibrium into the regions of nonlinearity. Further, the stock market returns show a more complicated distribution than a normal distribution [11]. They have sharper peaks and fat tails (Figure 1) [12].

Hence, there is a need to define a risk measure which is not bound by the constraints of CAPM. There have been several publications which argue that entropy is one such risk measure. In statistical mechanics, entropy is a measure of the number of unknown microscopic configurations of a thermodynamical system that is consistent with the measurable macroscopic quantities such as temperature, pressure, volume, etc. It is a measure of the uncertainty in the system [13][14]. In 1948, Shannon applied the concept of entropy as a measure of uncertainty in information theory, deriving Shannon entropy [15]. In finance, there are several features which make entropy more attractive as a risk measure. It is more general than the standard deviation [16][17] since it depends on the probabilities. Depending on the type of entropy used, it is capable of capturing the non-linearity in the dynamics of stock returns [18]. A review of applications of entropy in finance can be found in [19].

There have been several empirical studies comparing the predictive power of Rényi and Shannon entropies [20][21] with those from other measures (in particular $\beta$ and $\sigma$) with respect to portfolio expected returns. The conclusions are [21] that in the long run, the risk optimal portfolios from both Rényi and Shannon entropies show significantly lower variance than those from either $\sigma$ or $\beta$.

Several studies [22][23] indicate that the issues connected with the assumptions of CAPM (viz. efficient market hypothesis) can be addressed using statistical methods based on Tsallis entropy [24], which is a generalization of Shannon entropy to non-extensive systems. These methods were originally proposed to study classical and quantum chaos, physical systems far from equilibrium such as turbulent systems (non-linear), and long range interacting Hamiltonian systems. However, in the last several years, there has been considerable interest in applying these methods to analyze financial market dynamics as well. Such applications fall into the category of econophysics [25].

In an earlier work [26], we investigated the use of Tsallis relative entropy (TRE) [27], which is a generalization of Kullback-Leibler relative entropy (KLRE) [28] to non-extensive systems, as a new relative risk measure (relative to the market) for constructing portfolios that beat market returns. These investigations show that the relative risk – excess return profiles from TRE show a more consistent behavior than those from CAPM beta, both in terms of goodness of fit and the variation of returns with risk.

One aspect of concern in our earlier studies [26] is that the model distributions ($q$-Gaussian) of the stock market and equity returns estimated from the maximization of Tsallis entropy are symmetric in positive and negative returns. Observations of the data distributions show that they are not symmetric. In an earlier publication [29], we have shown that the model distributions of the financial market returns obtained from the non-extensivity parameters $q$ and the temperature parameters $b$ estimated taking asymmetry into account give a much better fit to the data distributions than the symmetric $q$-Gaussian distributions.

In the present work, we generalize the risk measure TRE to the asymmetric case (ATRE). It is assumed that both market and equity returns have asymmetric $q$-Gaussian distributions but with different model parameters. The model parameters are estimated using the procedure detailed in [29].

The rest of the paper is organized as follows. In Section 2, Tsallis relative entropy with some necessary background on Tsallis entropy and $q$-Gaussian distributions is discussed. Also discussed in this section is the generalization of TRE to the asymmetric case (ATRE). A relationship between ATRE and the parameters of asymmetric $q$-Gaussian distributions is derived. Section 3 deals with the data and methodology for constructing risk optimal portfolios using ATRE as the risk measure and the results. The conclusions are given in Section 4.

It should be noted that we use the terms volatility and risk interchangeably. Strictly speaking, the term volatility should be used since we only use the stock price time series for the analysis. However, in the literature the term risk has also been used to mean volatility.

The returns are calculated as defined in (1c). The term expected returns is used to mean predicted future average returns.

## 2. Theory

### 2.1 Review of Tsallis Statistics

Tsallis entropy is a generalization of Shannon entropy

$$S_{sh} = \sum_i P_i \, ln(1/P_i) \tag{2}$$

to non-extensive systems. It is given by

$$S_q = \sum_i P_i \, ln_q(1/P_i) \tag{3}$$

where $P_i$ is the probability density function at the i$^{th}$ sample under the condition $\sum_i P_i = 1$ and the $q$ logarithm $ln_q(x)$ is given by

$$ln_q(x) = (x^{1-q} - 1)/(1 - q) \tag{4}$$

The scaling parameter $q$ is a universal parameter, but its value can change from system to system.

Substituting (4) in (3), we get

$$S_q = \left(1 - \sum_i P_i^q\right)/(q - 1) \tag{5}$$

Unlike Shannon entropy, Tsallis entropy obeys a pseudo additive property

$$S_q(A + B) = S_q(A) + S_q(B) + (1 - q) \, S_q(A) \, S_q(B) \tag{6}$$

The scaling parameter $q$ denotes the extent of the non-extensivity of the system. As $q \to 1$, the additive property of Shannon entropy is recovered.

Considering the continuous case for a random variable $\Omega$, one can show [24] that the maximization of $S_q$ with respect to $P$ under the following constraints:

$$\int_{-\infty}^{\infty} P(\Omega)d\Omega = 1 \tag{7a}$$

$$\langle (\Omega - \overline{\Omega_q}) \rangle_q = \int_{-\infty}^{\infty} (\Omega - \overline{\Omega_q}) \, P^q(\Omega)d\Omega = 0 \tag{7b}$$

$$\langle (\Omega - \overline{\Omega_q})^2 \rangle_q = \int_{-\infty}^{\infty} (\Omega - \overline{\Omega_q})^2 P^q(\Omega)d\Omega \, / \int_{-\infty}^{\infty} P^q(\Omega)d\Omega = \sigma_q^2 \tag{7c}$$

gives the Tsallis $q$-Gaussian distribution

$$P_q(\Omega) = \frac{1}{Z_q} [1 + (q-1)b(\Omega - \overline{\Omega})^2]^{1/(1-q)} \tag{8}$$

$Z_q$ is the normalization and $b$ is the 'temperature parameter'. The expectation value $\langle - \rangle_q$ in (7b) and (7c) are the $q$–expectation values. $\overline{\Omega}$ is the mean value of $\Omega$.

Assuming that the variable $\Omega$ has zero mean, the normalization $Z_q$ is given by

$$Z_q = \int [1 + (q-1)b(\Omega)^2]^{1/(1-q)} \, d\Omega \tag{9a}$$

$$= C_q \, / \sqrt{b}$$

$$C_q = \sqrt{\pi} \, \frac{\Gamma\left(\frac{1}{q-1} - \frac{1}{2}\right)}{\sqrt{q-1} \, \Gamma\left(\frac{1}{q-1}\right)} \tag{9b}$$

Here $\Gamma$ is the gamma function. Note that in the limit $q \to 1$, it can be shown that the Tsallis entropy and the corresponding $q$-Gaussian distribution go to the Shannon entropy and the Gaussian distribution respectively.

## 2.2 Tsallis Relative Entropy

The generalization by Tsallis [27] of Kullback-Leibler relative entropy [28]

$$S_{KL}(P\|R) = -\sum_i P_i \, ln \, (R_i/P_i) \tag{10}$$

to non-extensive systems is given by

$$S_T(P\|R) = -\sum_i P_i \, ln_q(R_i/P_i) \tag{11}$$

P and R are normalized PDF's.

Using the definition of $ln_q(x)$ given in (4),

$$S_T(P\|R) = \left(\sum P_i(P_i/R_i)^{q-1} - 1\right)/(q-1) \tag{12}$$

The following are some of the properties of $S_T$ [30]:

1. Asymmetry: $S_T(P\|R) \neq S_T(R\|P)$                (13)

2. Non-negativity: Since $-ln_q(x)$ is a convex function for $q > 0$

$$S_T(P\|R) = -\sum_i P_i \, ln_q(R_i/P_i) \geq -ln_q\left(\sum_i P_i\left(R_i/P_i\right)\right) = 0 \tag{14}$$

3. Pseudo-additivity:

$$\begin{aligned} S_T(P_1 + P_2\|R_1 + R_2) = \; & S_T(P_1\|R_1) + S_T(P_2\|R_2) \\ & + (q-1)\, S_T(P_1\|R_1)\, S_T(P_2\|R_2) \end{aligned} \tag{15}$$

The first two properties hold for KL relative entropy as well.

Equation (15) shows the applicability of TRE to correlated systems. As $q \to 1$, the pseudo additivity becomes the additive property

$$S_{KL}(P_1 + P_2\|R_1 + R_2) = S_{KL}(P_1\|R_1) + S_{KL}(P_2\|R_2)$$

The analytical expressions for Tsallis relative entropy (TRE) $S_T$ in terms of the parameters of the $q$-Gaussian fit are given in [26].

## 2.3 Tsallis Relative Entropy from Asymmetric '$q$-Gaussian' distribution (ATRE)

As discussed in [26], direct calculations of relative entropies defined in equations (11) and (12) in terms of histograms of the data have several problems:

1. They depend on the number of bins in the histograms.
2. The relative entropies are defined only in the overlapping region of $R$ and $P$.
3. $S_{KL}$ is finite only if $R$ is non-zero in all the overlapping bins. The same is true for $S_q$ but for $q > 1$.

This makes the number of samples for the computation of relative entropies rather sparse and hence the stability and accuracy become questionable. However, if both $R$ and $P$ can be well fit with model distributions, analytical expressions for the relative entropies can be derived in terms of the parameters of these distributions. In [26] $R$ and $P$ were modelled with symmetric $q$-Gaussian distributions (8). Here it was assumed the individual equity distributions $P$ have the same non-extensivity parameter $q$ but different temperature parameters $b$. The analytical formulas for TRE obtained were used as relative risk measures to investigate risk-return relationships.

To generalize the risk measure TRE to ATRE, we follow a procedure as discussed in [29]. Assuming $R$ and $P$ have zero mean, their corresponding distributions are given by

$$R(\Omega) = [R_-(\Omega) + R_+(\Omega)] \tag{16}$$

where

$$R_-(\Omega) = 0 \qquad \Omega > 0 \tag{17a}$$

$$R_+(\Omega) = 0 \qquad \Omega \leq 0 \tag{17b}$$

$R_-(\Omega)$ and $R_+(\Omega)$ are half $q$-Gaussians, given by

$$R_-(\Omega) = \frac{1}{Z_-}\left[1 + (q_- - 1)b_-\Omega^2\right]^{1/(1-q_-)} \qquad \Omega \leq 0 \tag{18a}$$

$$R_+(\Omega) = \frac{1}{Z_+}\left[1 + (q_+ - 1)b_+\Omega^2\right]^{1/(1-q_+)} \qquad \Omega > 0 \tag{18b}$$

Here, $q_-$, $b_-$, $q_+$ and $b_+$ are the $q$-Gaussian parameters for negative and positive $\Omega$ respectively. The normalizations $Z_-$ and $Z_+$ are such that

$$\int_{-\infty}^{0} R_-(\Omega)d\Omega = \int_{0}^{\infty} R_+(\Omega)d\Omega = \frac{1}{2} \tag{19}$$

so that the complete PDF $R(\Omega)$ in (16) is normalized. This gives

$$Z_- = C_{q_-}/(\sqrt{b_-}) \tag{20a}$$

$$Z_+ = C_{q_+}/(\sqrt{b_+}) \tag{20b}$$

The $C_q$'s are as defined in (9).

The distributions $P$ for individual equities are obtained by replacing $R$, $q$ and $b$ by $P$, $q'$ and $b'$ respectively in equations (16) – (20b).

The procedure for estimating the parameters $q$ and $b$ for asymmetric $q$-Gaussian distributions is described in detail in [29].

For the asymmetric case, the non-extensivity parameter $q$ in the definition of Tsallis relative entropy given in (11) is dependent on the sign of the random variable. Hence the relative entropy (ATRE) in this case becomes

$$S_T^{(a)}(P \| R) = [S_- + S_+] \tag{21a}$$

where

$$S_- = -\sum_{i_-} P_{i_-} \, ln_{q_-}(R_{i_-}/P_{i_-}) \tag{21b}$$

$$S_+ = -\sum_{i_+} P_{i_+} \, ln_{q_+}(R_{i_+}/P_{i_+}) \tag{21c}$$

Here $i_-$ and $i_+$ represent $\Omega < 0$ and $\Omega > 0$ for the discrete case.

For brevity, we write S to represent either $S_-$ or $S_+$. To express S in terms of the model parameters, we define new parameters

$$\varphi = 1/(q-1) \text{ and } \kappa = (q-1)b \tag{22a}$$

$$\gamma = \varphi'/\varphi \qquad \text{and } \eta = \sqrt{\kappa / \kappa'} \tag{22b}$$

As defined earlier, the primed quantities are the parameters for the distribution $P$.

As shown in the Appendix, for $q, q' > 1$

$$S = \frac{\varphi}{2}\left[ (N-1) + N\left(\frac{1}{2}\eta^2 / \left(\gamma + \varphi' - \frac{3}{2}\right)\right) \right] \tag{23}$$

where

$$N = \left(Z/Z'\right)^{1/\varphi} \left(B(\gamma, \varphi')/B\left(\gamma, \varphi' - \frac{1}{2}\right)\right) \tag{24}$$

$Z, Z'$ are the same as (9a) and (9b), expressed in terms of $\varphi, \kappa$ and $\varphi', \kappa'$ respectively.

$B$ is the Beta function.

$S_-$ and $S_+$ are obtained by replacing $q$, $q'$, $b$, $b'$ in (22a) - (24) with $q_-$, $q'_-$, $b_-$, $b'_-$ and $q_+$, $q'_+$, $b_+$, $b'_+$ respectively.

With the assumption $q = q'$, $(S_- + S_+)$ goes over to the symmetric case discussed in [26]

$$S_T(P\|R) = \left[ -ln_q(\eta) + \tfrac{1}{2}\eta^{(1-q)}(\eta^2 - 1) \right] \tag{25}$$

## 3. Data, Methodology, and Results

### 3.1 Data

For the present study, we consider daily stock data from 6 March 1995 to 24 November 2021 (6688 samples). The reference market index is chosen to be SPY. This is an ETF which closely follows the S&P 500 index. The data are adjusted for dividends and splits. No attempt has been made to correct the data for inflation. Note that this data covers the .com bubble, 2008 crash, and the COVID periods.

### 3.2 Methodology

Computation of the relative entropy (risk measure) given in (23) involves estimating the parameters $\{q_-,\ q_+,\ b_-,\ b_+\}$ and $\{q'_-,\ q'_+,\ b'_-,\ b'_+\}$ for the reference index and individual equities respectively. For estimating these, we follow the procedure described in [29].

In testing the performance of the risk measures in this study, we follow a procedure somewhat similar to that described by Black, Jensen and Scholes [10]. The exact procedure is as follows.

For each cycle, we use the list of securities in the SPY ETF as of November 2021 which have data extending five years before the start date of the cycle, e.g., all the way back to March 1995 for the first cycle. As the cycles move forward in time, more and more securities enter the computation. This gives us about 340 stocks in the first cycle increasing to about 460 in the last cycle.

Each cycle consists of the following three steps:

a) About five years of data (1400 samples to be exact) prior to the starting date of the cycle are used to estimate the parameters of the risk models for the reference market index SPY and for each of the securities in it. That makes 18 September 2000 the starting date for our first cycle. From these prior samples, ten day percent returns (1c) are calculated and used to estimate the model parameters. These are then used to calculate the relative entropy risk measures (23). The expected return in excess of market return is computed as the six month percent return (a typical portfolio turnover time) from the data six months in future.

b) The risk values are then binned and the securities are assigned to each bin according to their risk value. The set of securities in each bin can be considered a portfolio with the center of each bin as the risk value of the portfolio. The bin widths are chosen from the minimum and maximum values of the risk measure in the first cycle such that there is, initially, a given number of securities in each bin. Two additional bins are added at the high end for the possibility of higher risk values in future cycles. The number of bins and bin widths are kept fixed. Hence the number of securities in each bin (portfolio) can change as time proceeds. In order to understand the effect of asymmetry on the earnings, the binnings are done separately for the risks $S_-$, $S_+$, and the total $[S_- + S_+]$.

c) Assuming an equal amount of money invested in every security, the expected return of the portfolio in each bin in excess of the SPY return is calculated. This gives the risk-return values for each portfolio.

The data are then shifted by six months and steps a) - c) are repeated for the next cycle.

Finally, expected portfolio returns in each bin are averaged over all cycles to get the mean risk-excess return ($E_{rel}$). With our present data, there are 42 six month cycles.

The choice of 1400 samples window size for estimating parameters was made based on following considerations.

This window size gives roughly comparable number of samples for both positive and negative returns. This is essential to ensure that the fit is good for both the positive and negative return branches of the distribution. Smaller window sizes do not always ensure this, which would result in higher errors in the estimated parameters.

In general, larger windows are better if there is enough data. But in the present case, quite a few stocks do not have electronic data available before 1995. So with larger windows, the statistics in terms of number of stocks gets poorer. In addition, larger windows require starting at a later date, which reduces the number of cycles, again resulting in poorer statistics.

Note that we have used 10 day returns for the estimation of parameters. One day returns are very noisy and tend to have sharp spikes. For longer delays, the $q$ parameter, especially for positive returns, tends to 1 [29]. Hence a delay of 10 days seems to be reasonable both in terms of reducing noise and retaining the non-extensive character of returns.

It should also be noted that every time the data are shifted, the contents of each bin in step b) can change. Also, in step c), each bin is rebalanced every six months such that an equal amount of money is invested in every security. This means that if this procedure is applied in practice, some securities would be sold and others bought every six months to implement steps b) and c). The effects on the portfolio returns due to transaction costs incurred in such selling and buying and taxes imposed on realized gains are not included in this study.

### 3.3 Goodness of Fit

The portfolio performance using the ATRE as the risk measure depends on:

(a) how well our model distributions fit the data distributions
(b) how close the risk-return patterns are to a linear regression

Figure 2 shows a comparison of the asymmetric $q$-Gaussian distributions with the data distributions for the reference stock SPY and a few randomly chosen equities from its constituents. Visual inspection shows that the fits are pretty good. However, to quantify the 'goodness of fit,' Kolmogorov-Smirnov (KS) [31] tests are carried out. Briefly, this involves determining the maximum absolute distances $\mathbf{D_{max}}$ between the empirical and the synthetic $q$-Gaussian cumulative distribution functions (CDF). The fit is good if $\mathbf{D_{max}}$ is less than a critical distance $\mathbf{D_{crit}}$. The details of constructing synthetic $q$-Gaussians and determining $\mathbf{D_{max}}$ and $\mathbf{D_{crit}}$ are given in [26]. For all the stocks displayed in Figure 2, $\mathbf{D_{max}}$ is ~0.1 and $\mathbf{D_{crit}}$ ~0.3 showing that the distributions of the returns of even the individual stocks can be modelled well with asymmetric $q$-Gaussian distributions.

To assess how well linear regression works for the risk-return patterns, we estimate $\chi^2$, which is one of the commonly used estimates in statistics [21] in determining the goodness of fit. This quantity shows how close the risk-return return relationships are to a linear fit. If $\{s\}$ is a set of risk values of the bins and $\{e\}$ the corresponding portfolio earnings, then

$$\chi^2 = 1 - \frac{\sum_i [e_i - (p_0 + p_1 s_i)]^2}{\sum_i (e_i - \bar{e})^2} \tag{26}$$

Here $p_0$ and $p_1$ (intercept and slope) are the parameters of the linear fit and $\bar{e}$ is the mean of $e$. Note that the closer the values of $e$ to the linear fit, the closer is $\chi^2$ to 1.

### 3.4 Results

Figures 3 and 4 show the long term behavior of the $E_{rel}$ vs. the risk ATRE (equation 21a) and the TRE (symmetric case). The period is 2000-2021. Also shown are the linear fits and the goodness of the fits $\chi^2$. Note that for this long period, the slopes of the linear fit in both the cases are positive, indicating that for greater relative risk there is greater relative return. This behavior is like that observed in the tests of the CAPM model [10] as well. The $\chi^2$ is good and comparable in both the cases.

In Figures 3 and 4, the maximum risk value is cut off at 2. This is because, beyond this value, either there is not enough multiplicity in the bins or the linear fit is considerably degraded.

Tests of CAPM by Black, Jensen and Scholes [10] for shorter periods (9-10 years) show that the linear relationship between risk and return is intrinsic and not the result of better statistics. However, the risk-return patterns are non-stationary, i.e., the slopes and intercepts vary widely for each period, the slopes becoming even negative in some cases. Here we carry out similar tests for both the symmetric and non-symmetric cases. We first divide the data interval into two periods of 10 years each: a) 18 September 2000 – 27 September 2010 and b) 28 March 2011 – 8 June 2021. The first interval covers the .com bubble and part of the 2008 crash. The second interval covers both the crash and the COVID periods.

Figures 5 and 6 show the 10 year behavior of the $E_{rel}$ vs. the relative risk measures for the asymmetric case (ATRE) and the symmetric case respectively. Note that even for this shorter term, the slopes of the linear fits are positive, indicating that for greater relative risk there is greater relative return. The risk-return patterns for the asymmetric and symmetric cases are very similar.

In Figures 3 and 5 the risk measure plotted is the total entropy $(S_- + S_+)$. To understand the effect of asymmetry on risk-return patterns, we need to look at the effect of $S_-$ and $S_+$ separately. In [29] we showed that, as a function of time scale, the non-extensivity is much more pronounced for negative returns than for positive returns. Figure 7 shows the dynamical variations of $q_-$ and $q_+$ every 6 months from $2000 - 2021$ for the reference stock SPY. For the periods covering the 2008 crash and COVID, the $q_-$ values are significantly higher than $q_+$ indicating that the negative returns data have higher non-extensive character than the positive returns. Figures 8 and 9 show the risk-return profiles for $S_+$ and $S_-$ separately. The earning behavior in the case of $S_-$ for the periods which include the 2008 crash and COVID (Figure 9b) is very different from all the others (Figures $3 - 8$). The slope is much higher than the other slopes (higher earnings for the same risk value). This clearly indicates the non-stationarity character of the risk-return patterns when $S_-$ is used as the risk measure.

As seen above, portfolio earnings increase with risk and the highest risk bin with reasonable multiplicity gives highest earnings. However, because of non-stationarity, the risk range can change with time. Figure 10 shows the monthly variation of the 90th percentile risk value (i.e., 90% of the equities have risks below this value) as a function of time for three risk measures: (a) the total entropy for the asymmetric case, (b) the entropy for symmetric case, and (c) $S_-$. All three cases show non-stationarity, with the first two cases showing a much higher degree of variation than $S_-$. Hence, if a fixed risk value is chosen over the life of portfolio, particularly a high risk value, one has to make sure that it has reasonable multiplicity of stocks to construct the portfolio during the turnover times.

Figure 11 shows the variation of 10 year cumulative earnings at maturity, as a function of the start time of the portfolio. The risk measures used are the 90th percentile relative entropy values. The turnover time for the portfolios is six months. The start time shifts by one month. The portfolio consists of 15 stocks around the 90th percentile risk value. All three portfolios yield earnings more than that from SPY for all investment start times.

Let us now look at the 10 year earnings when a fixed risk is chosen over the life of the portfolio. To ensure enough multiplicity, we look at the low risk situation by choosing the minimum value of the $90^{th}$ percentile variation (Figure10). This is displayed in Figure 12. Note that the portfolio earnings for the risk type $S_-$ stays above SPY earnings for all periods investigated. This however is not true for the asymmetric and symmetric risk types.

The statistics of the cumulative earning series shown in Figures 11 and 12, corresponding to the $90^{th}$ percentile and a fixed low value of the risk measures, are given in Tables 1 and 2, respectively. As expected, the earnings are lower for the fixed low risk case than those for the $90^{th}$ percentile case where the risks can vary from high to low values.

## 4. Summary and Conclusions

In this work, we have considered the asymmetric character of the market and equity returns distributions in defining a new risk measure for the selection of risk optimal portfolios whose returns, in the long term, are expected to exceed market returns. The new risk measure (ATRE) is the generalization of Tsallis relative entropy (TRE), which was investigated earlier [26], as the risk measure. The distributions of the returns of both the market (SPY S&P 500 ETF) and the individual stocks can be well fit with a distribution which is a linear combination of two half $q$-Gaussian distributions which have different non-extensivity parameter $q$ and temperature parameter $b$. The relative entropy ATRE can be analytically expressed in terms of these model parameters. This alleviates several problems (described in section 2.3) encountered in the histogram based estimation of relative entropies.

The present empirical tests show that the negative returns exhibit a much more pronounced non-extensive character than the positive returns, especially during the crash and COVID periods. This is indicated by the higher $q_-$ values. A comparison of the risk-excess return profiles of ATRE with TRE and $S_-$ show that all the three profiles have positive slopes (average earnings increase with risk). This is true both for long term (~20 years) and short term (~10 years) portfolios. Over the shorter term, consisting of periods of very different market characteristics (bubble, crash and COVID), the measures ATRE and TRE still show similar behavior in terms of slopes and goodness of fit ($\chi^2$). This, however, is not the case for $S_-$. The 10 year profiles generated from the data that include the .com bubble are very different from those generated during the chaotic periods of crash and COVID. The latter has a much higher slopes indicating higher returns. This shows that with a proper risk measure, higher portfolio earnings can be expected even if the market goes through chaotic periods during the lifetime of the portfolio.

The non-stationary behavior of the 10 year cumulative portfolio earnings as a function of the starting time of the portfolio shows that during certain periods the earnings can come very close to or even dip below the market earnings (zero or negative excess return). This is particularly the case with relatively low risk portfolios using TRE and ATRE as the risk measures. However, in the case of $S_-$, the excess portfolio earning stays positive for all start times, irrespective of market behavior.

The empirical investigations in this work point to the importance of taking into account the asymmetry and non-extensivity of the financial markets in defining risk measures. $S_-$ is one such measure and may help in the construction of portfolios whose returns beat the markets even when the latter goes through chaotic situations.

## Appendix: Derivation of Asymmetric Tsallis Relative Entropy (ATRE)

### A.1 Asymmetric Case

Denoting

$$\varphi = 1/(q-1) \ \text{ and } \ \kappa = (q-1)b \tag{A1}$$

Since $\varphi$ now depends on the sign of $\Omega$, the integral representation of Tsallis relative entropy (12) can be re-written as

$$S_T(P\|R) = \int_{-\infty}^{\infty} \varphi\left\{P(P/R)^{1/\varphi} - P\right\} d\Omega \tag{A2}$$

Breaking the integral into negative and positive $\Omega$, (A2) can be written as

$$S_T(P\|R) = \varphi_- \left\{\int_0^\infty P_-(P_-/R_-)^{1/\varphi_-} d\Omega_- - \frac{1}{2}\right\}$$
$$+ \ \varphi_+ \left\{\int_0^\infty P_+(P_+/R_+)^{1/\varphi_+} d\Omega_+ - \frac{1}{2}\right\} \tag{A3}$$

Here the $-, +$ signs stand for negative and positive returns respectively.

For brevity, we drop the suffixes $-, +$ and denote

$$S \ = \varphi\left\{\int_0^\infty P(P/R)^{1/\varphi} d\Omega \ - \frac{1}{2}\right\} \tag{A4}$$

Bear in mind that $\varphi$ and $\Omega$ stand for either $\varphi_-, \ \Omega_-$ or $\varphi_+, \ \Omega_+$.

Using the model distributions $(16 - 20b)$

$$R(\Omega) = \frac{1}{Z}\left[1 + \kappa \ \Omega^2\right]^{-\varphi} \tag{A5}$$

$$P(\Omega) = \frac{1}{Z'}\left[1 + \kappa'\Omega^2\right]^{-\varphi'} \tag{A6}$$

Remember from section 2.3 that the unprimed quantities refer to the model for the reference index and the primed quantities refer to the models for the individual equities.

Substituting (A5) and (A6) in (A4)

$$S = \varphi \left[ (Z/Z')^{1/\varphi} \frac{1}{Z'} \left\{ \int_0^\infty (1 + \kappa\Omega^2)/(1 + \kappa'\Omega^2)^{\gamma(1+\varphi)} d\Omega \right\} - \frac{1}{2} \right] \quad \text{(A7)}$$

where $\gamma = \varphi'/\varphi$.

Using the following integration results with $\eta = \sqrt{\kappa / \kappa'}$ and where $B$ is the Beta function

$$\frac{1}{Z'} \left\{ \int_0^\infty 1/(1 + \kappa'\Omega^2)^{\gamma(1+\varphi)} d\Omega \right\} = \frac{1}{2} \left( B(\gamma, \varphi')/B\left(\gamma, \varphi' - \frac{1}{2}\right) \right) \quad \text{(A8)}$$

and

$$\frac{1}{Z'} \left\{ \int_0^\infty \kappa'\Omega^2/(1 + \kappa'\Omega^2)^{\gamma(1+\varphi)} d\Omega \right\} =$$

$$\left( \frac{1}{4}\eta^2 / \left(\gamma + \varphi' - \frac{3}{2}\right) \right) \left( B(\gamma, \varphi')/B\left(\gamma, \varphi' - \frac{1}{2}\right) \right) \quad \text{(A9)}$$

Substituting (A8) and (A9) in (A7)

$$S = \frac{\varphi}{2} \left[ (N - 1) + N \left( \frac{1}{2}\eta^2 / \left(\gamma + \varphi' - \frac{3}{2}\right) \right) \right] \quad \text{(A10)}$$

where

$$N = \left( \frac{Z}{Z'} \right)^{1/\varphi} \left( B(\gamma, \varphi')/B\left(\gamma, \varphi' - \frac{1}{2}\right) \right) \quad \text{(A11)}$$

Therefore, from (A3), ATRE is

$$S_T(P\|R) = S_- + S_+ \quad \text{(A12)}$$

where $S_-$ and $S_+$ are obtained by replacing $\varphi, \kappa, \varphi', \kappa'$ by $\varphi_-, \kappa_-, \varphi'_-, \kappa'_-$ and $\varphi_+, \kappa_+, \varphi'_+, \kappa'_+$ respectively in (A10).

## A.2 Symmetric Case

We will now show that (A12) goes over to the symmetric case discussed in [26] where it is also assumed $q = q'$. In this case

$$S_- = S_+$$

$$\frac{Z}{Z'} = \sqrt{(\kappa'/\kappa)} = 1/\eta$$

$$B(\gamma, \varphi') \,/\, B(\gamma, \varphi' - 1/2) \;=\; (\varphi - 1/2)/\varphi$$

$$N = \sqrt{\eta}^{\,-(1/\varphi)} (\varphi - 1/2)/\varphi$$

Using the transformations (A1) and re-arranging terms, it is straight forward to show that

$$S_T(P\|R) = \left[ -ln_q(\eta) + \frac{1}{2}\eta^{(1-q)}(\eta^2 - 1) \right] \qquad \text{(A13)}$$

This is the same as equation (20) in [26] for standardized returns.

**Figures**

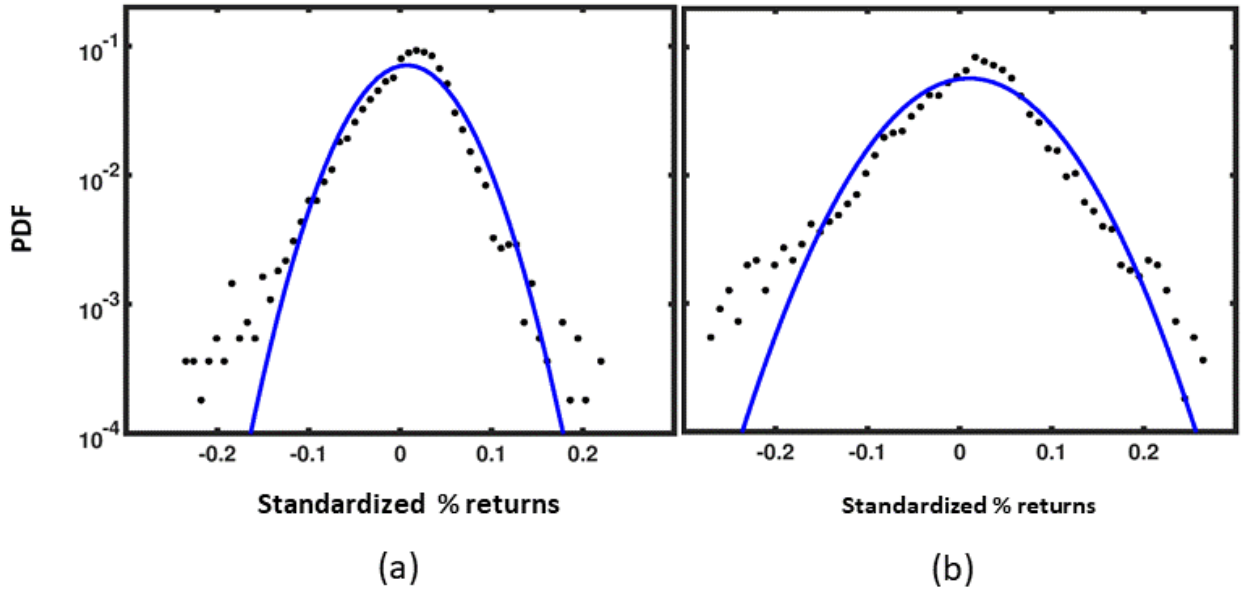

(a)                                           (b)

Figure 1. Comparison of the distributions of monthly standardized percent returns with the Gaussian distributions (solid blue line) having the same mean and standard deviation as the data (black dots). (a) S&P 500 for the period January 1995 – January 2017 and (b) Nasdaq over the same period.

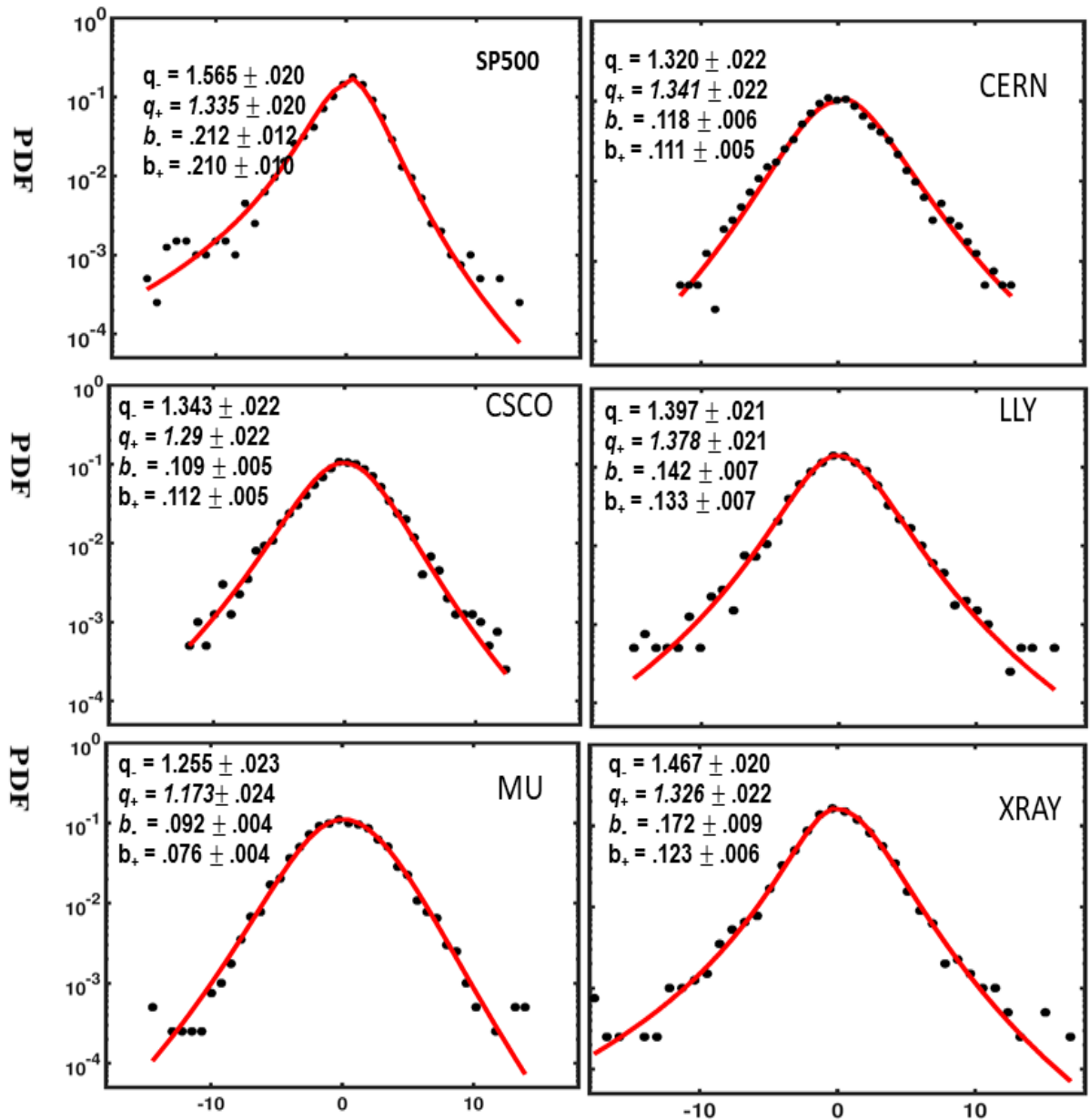

Figure 2. *q*-Gaussian fit to the distributions of ten day standardized percent returns of SPY (S&P 500 ETF) and five randomly chosen stocks from the SPY list of stocks. The ticker symbols of the stocks are displayed on each corresponding figure. The estimated *q*-Gaussian parameters and their corresponding errors are also shown. Period March 1995 – January 2022.

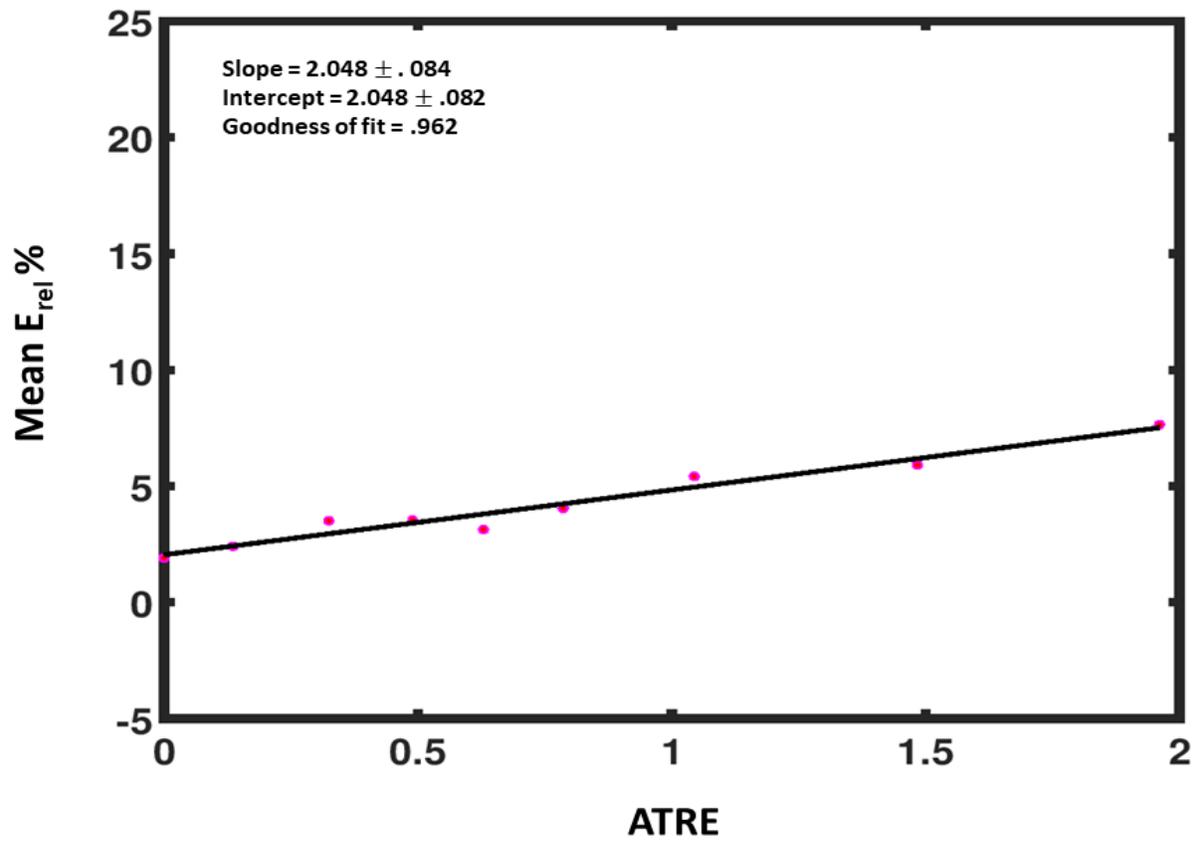

Figure 3. Average six month excess returns of the portfolios vs. risk measure $(S_- + S_+)$ for the asymmetric case (ATRE). Data (earnings) interval 2000-2021.

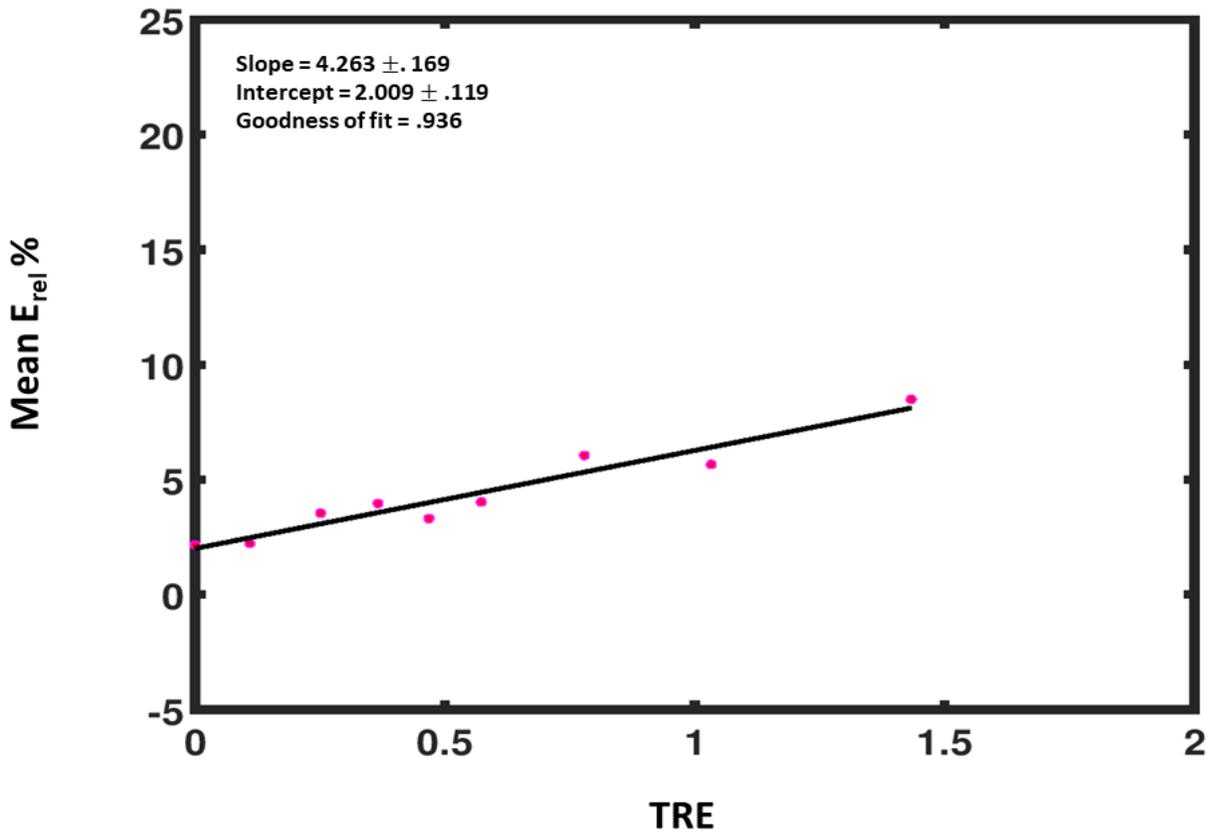

Figure 4. Same as Figure 3 for the symmetric case (TRE).

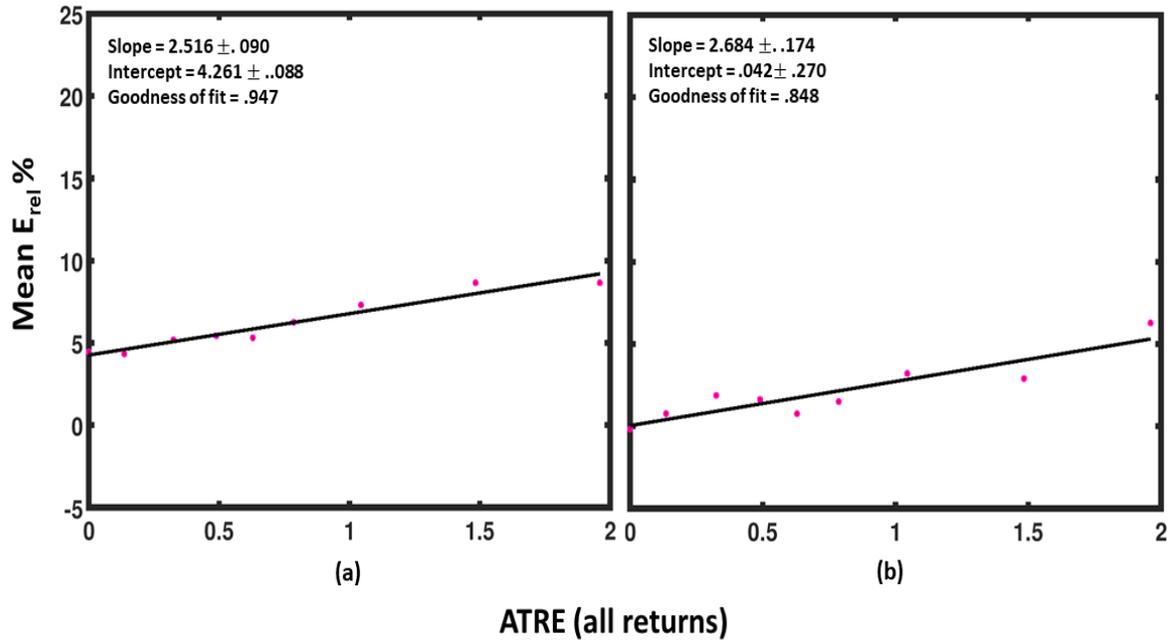

Figure 5. Average six month excess returns of the portfolios vs. risk measure $(S_- + S_+)$. (a) for the 10 year period 18 September 2000 – 27 September 2010 and (b) same for 28 March 2011 – 8 June 2021.

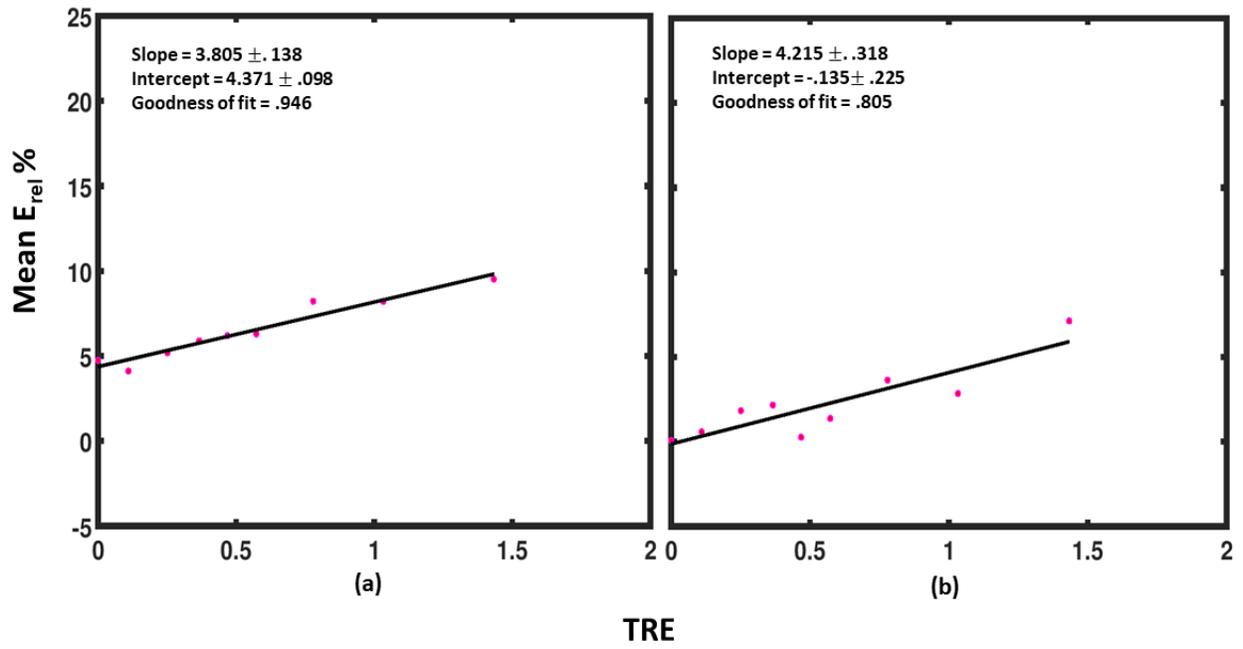

Figure 6. Same as Figure 5 for the symmetric case.

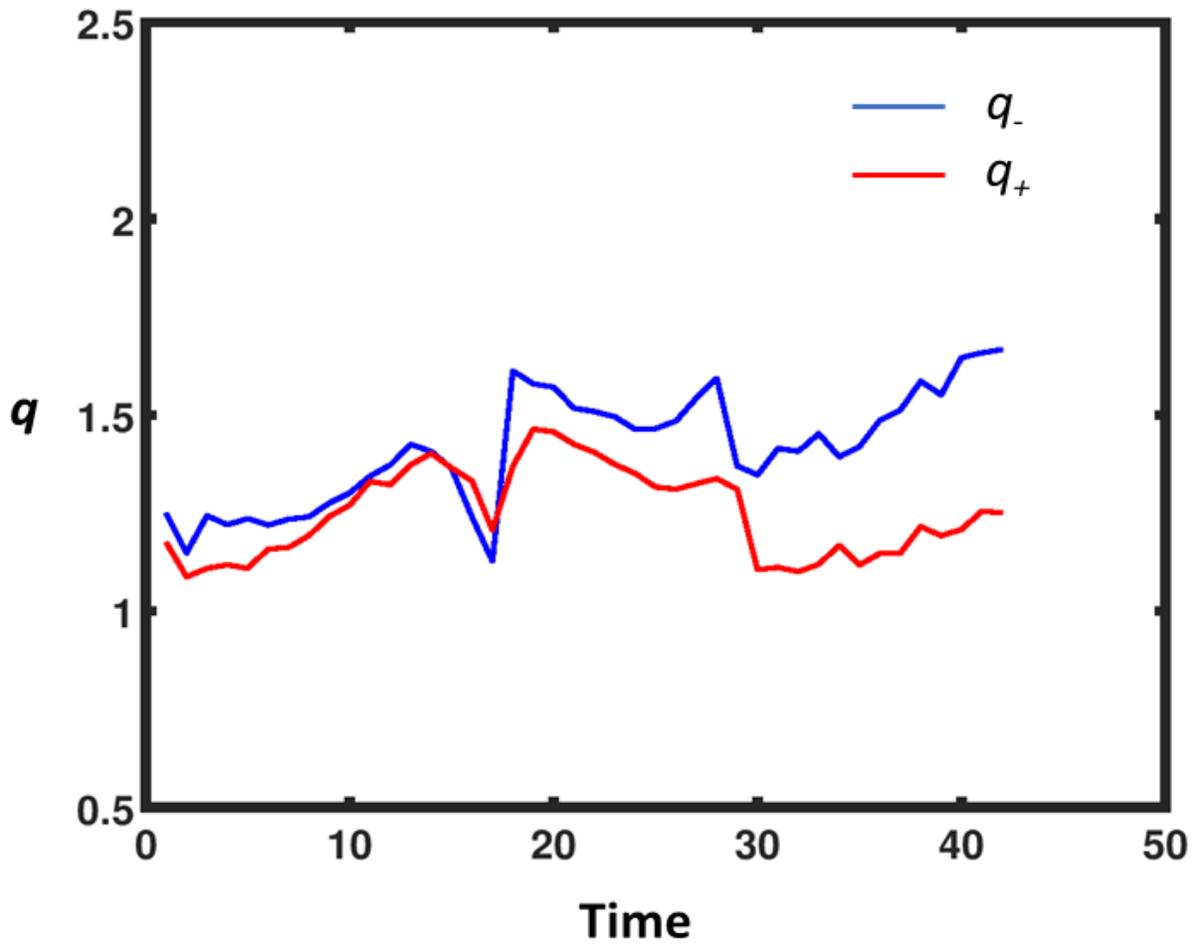

Figure 7. Variation of $q_-$ and $q_+$ with time over the period 18 September 2000 – 24 November 2021. The time sample interval is 6 months (same as turn over time).

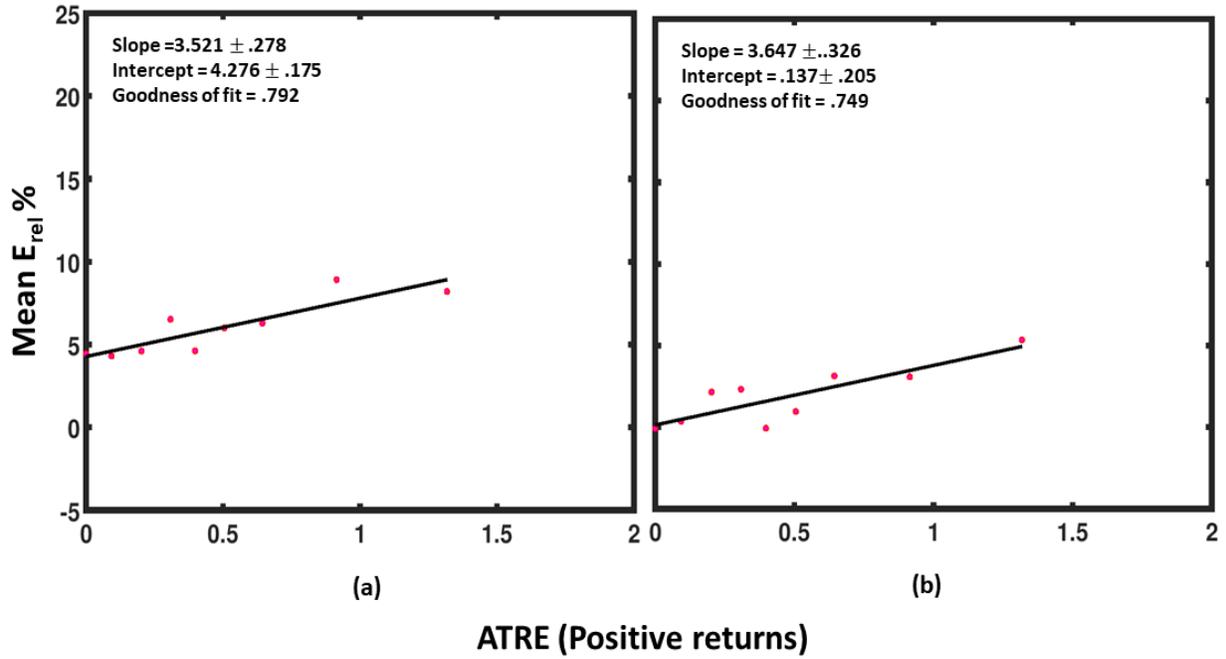

Figure 8. Same as Figure 5 with $S_+$ as the risk measure.

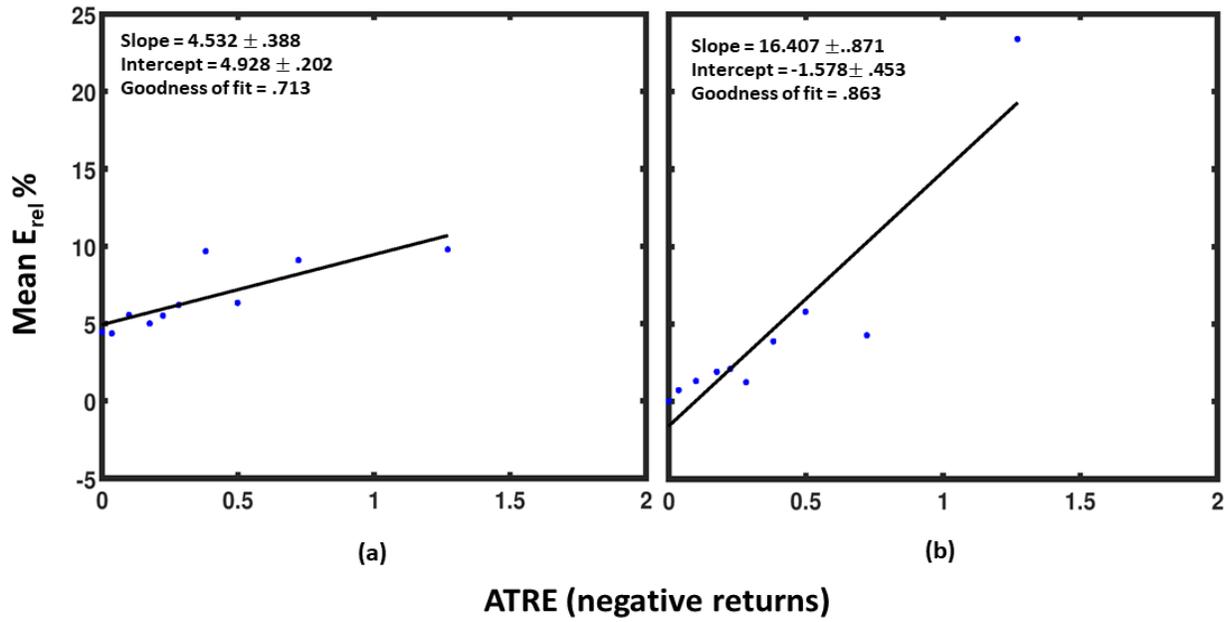

Figure 9. Same as Figure 5 with $S_-$ as the risk measure.

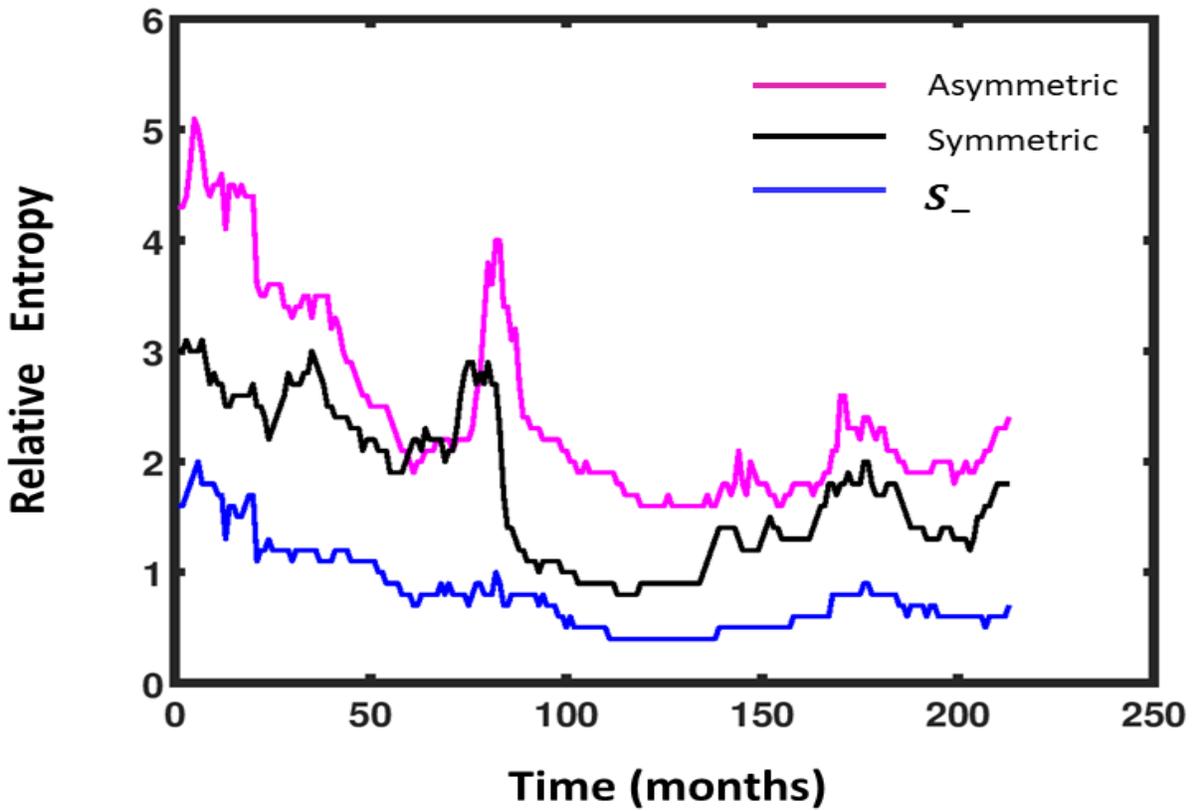

Figure 10. Variation of relative entropies (risk measures) with time over the period 18 September 2000 – 24 November 2021. The time sample interval is one month.

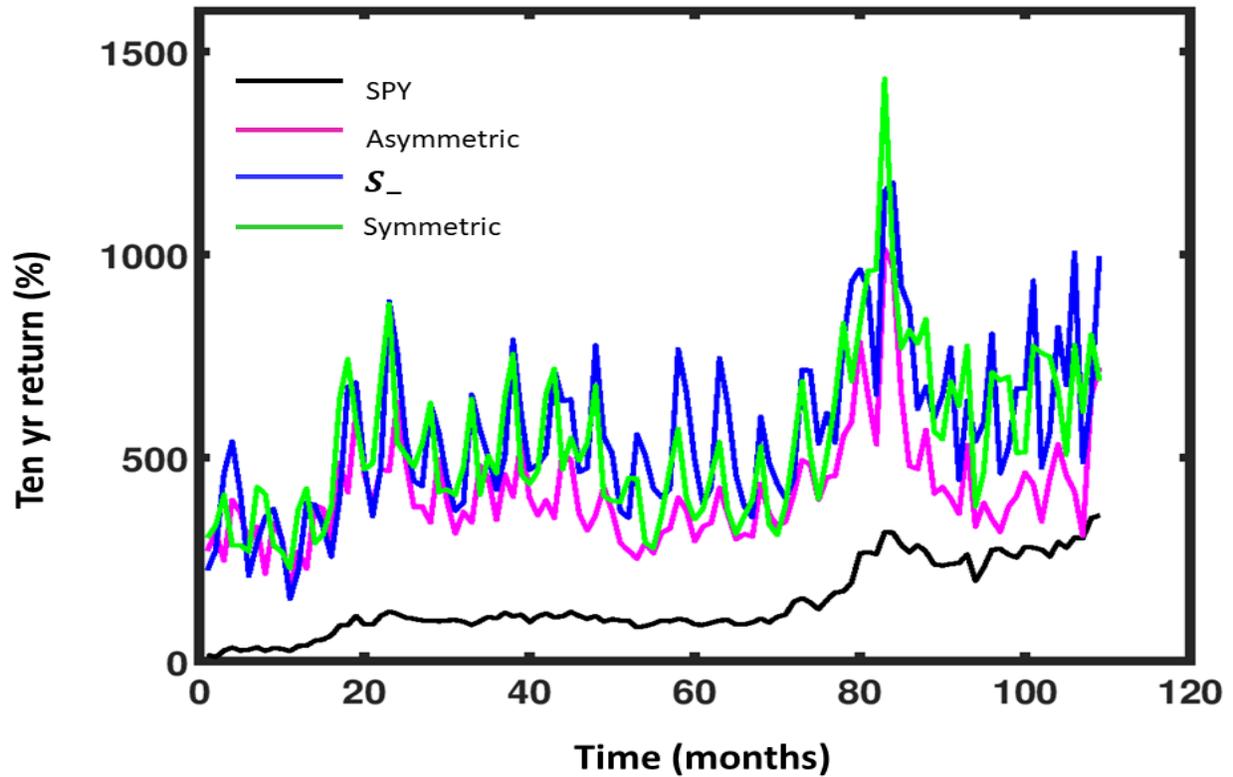

Figure 11. Ten year SPY and portfolio earnings as a function of the start time of portfolio. The x-axis is the maturity date. Time digitization is 1 month. The risks used are the 90[th] percentile values. The first 10 year portfolio maturity date is 2 December 2010 and the last 27 October 2021.

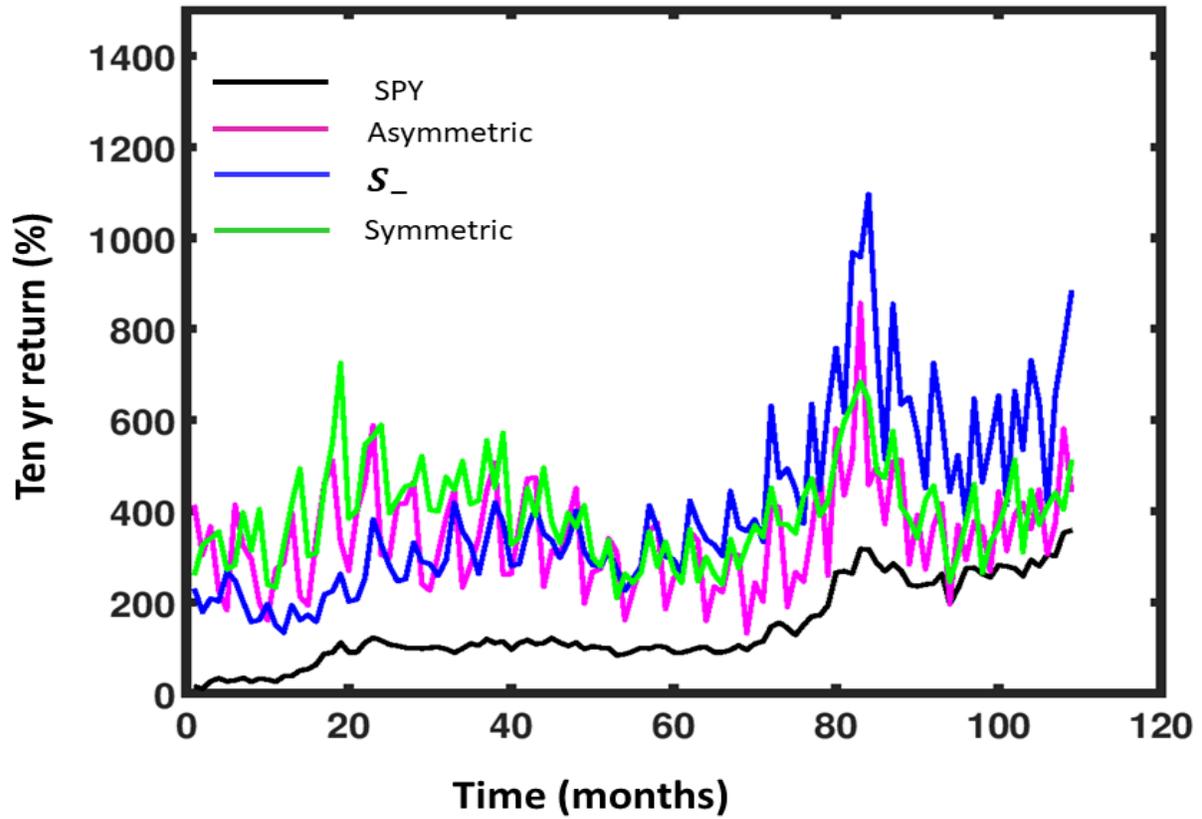

Figure 12. Same as Figure 11, but for fixed risk values over 10 years (maturity date for the portfolio for a given starting time). Asymmetric risk of 1.9, $S_-$ risk of 0.6, Symmetric risk of 1.0.

| Relative Entropy (risk) | % Earnings (Mean) | % Earnings (Median) | Standard Deviation |
|---|---|---|---|
| Asymmetric ($S_- + S_+$) | 416% | 389% | 134 |
| Asymmetric ($S_-$) | 565% | 535% | 201 |
| Symmetric | 542% | 507% | 197 |
| SPY ETF | 144% | 108% | 89 |

Table 1. The earning statistics for 10 year portfolios shown in Figure 11 (90th percentile risk).

| Relative Entropy (risk) | % Earnings (Mean) | % Earnings (Median) | Standard Deviation |
|---|---|---|---|
| Asymmetric ($S_- + S_+$) | 344% | 336% | 115 |
| Asymmetric ($S_-$) | 404% | 351% | 199 |
| Symmetric | 396% | 391% | 106 |
| SPY ETF | 144% | 108% | 89 |

Table 2. The earning statistics for 10 year portfolios shown in Figure 12 (fixed risk).